\documentclass[useAMS,usenatbib]{mn2e}

\usepackage{epsfig}
\usepackage{myaasmacros}

\begin{document}

%% LaTeX will automatically break titles if they run longer than
%% one line. However, you may use \\ to force a line break if
%% you desire.

\title[Specific Angular Momentum of Disc Merger Remnants and the $\lambda_R$-Parameter]{Specific Angular 
Momentum of Disc Merger Remnants and the $\lambda_R$-Parameter}

%% Use \author, \affil, and the \and command to format
%% author and affiliation information.
%% Note that \email has replaced the old \authoremail command
%% from AASTeX v4.0. You can use \email to mark an email address
%% anywhere in the paper, not just in the front matter.
%% As in the title, use \\ to force line breaks.

\author[R. Jesseit, M. Cappellari, T. Naab, E. Emsellem  and A. Burkert]
{Roland Jesseit $^{1}$\thanks{E-mail: jesseit@usm.uni-muenchen.de}, Michele Cappellari$^{2}$, Thorsten
  Naab$^{1}$, Eric Emsellem$^{3}$  \newauthor and Andreas Burkert$^{1}$\\
$^1$ Universit\"atssternwarte M\"unchen, Scheinerstr 1, 81679 M\"unchen, Germany \\
$^2$ Sub-Department of Astrophysics, University of Oxford, Denys Wilkinson
Building, Keble Road, Oxford, OX1 3RH, UK\\
$^3$ Universit\'e de Lyon, Lyon, F-69003, France ; Universit\'e Lyon~1,
Observatoire de Lyon, 9 avenue Charles Andr\'e, Saint-Genis
Laval, \\ F-69230, France ;
CNRS, UMR 5574, Centre de Recherche Astrophysique
de Lyon ; Ecole Normale Sup\'erieure de Lyon,\\ Lyon, F-69007, France\\}
\date{Draft version. Accepted ??ccn?. Received ??? in original form ???}
\pagerange{\pageref{firstpage}--\pageref{lastpage}} \pubyear{2002}
\maketitle
\label{firstpage}

\begin{abstract}
We use two-dimensional kinematic maps of simulated binary disc mergers to investigate  
the $\lambda_R$-parameter, which is a luminosity weighted measure 
of projected angular momentum per unit mass. This parameter was introduced to subdivide the 
SAURON sample of early type galaxies in so called fast $\lambda_R > 0.1$ and slow 
rotators $\lambda_R < 0.1$. Tests on merger remnants reveal that $\lambda_R$ is a robust indicator 
of the true angular momentum content in elliptical galaxies. We find the same range 
of $\lambda_R$ values in our merger remnants as in the SAURON galaxies. 
The merger mass ratio is decisive in creating a slow or a fast rotator in a single binary merger, the 
former being created mostly in an equal mass merger. Slow rotators have a $\lambda_R$ which does not 
vary with projection. The confusion rate with face-on fast rotators is very small. Merger with low 
gas fractions form slow rotators with smaller ellipticities and are in much better agreement with the 
SAURON slow rotators. Remergers of merger remnants are slow rotators but tend to have too high 
ellipticities. Fast rotators maintain the angular momentum content from the progenitor disc galaxy 
if merger mass ratio is high. Some SAURON galaxies have values of $\lambda_R$ as high as our 
progenitor disc galaxies.

\end{abstract}

%% Keywords should appear after the \end{abstract} command. The uncommented
%% example has been keyed in ApJ style. See the instructions to authors
%% for the journal to which you are submitting your paper to determine
%% what keyword punctuation is appropriate.

\section{Introduction}
The connection between projected and the intrinsic properties of elliptical galaxies is a notorious problem. 
The Hubble classification of early-type galaxies is based on their elliptical shape, which was the natural observable
distinctive feature. It was already pointed out by Hubble that this sequence is only a sequence of apparent shape
depending on the inclination under which we observe a galaxy. When kinematic observations of ellipticals became available,
they showed that some of them rotate too slowly to be shaped by 
rotation alone \citep{1977ApJ...218L..43I}. 
\citet{1978MNRAS.183..501B} proposed that these systems are probably shaped to a large
extend by anisotropic pressure. This is best seen when the apparent 
flattening is plotted versus the balance between ordered and unordered motion $(v/\sigma, \epsilon)$ using the relation of 
these observables for an isotropic rotator as a reference point. Observations also showed that the amount of 
rotation varies with luminosity 
\citep{1983ApJ...266...41D}  and isophotal shape \citep{1988A&A...193L...7B}. Isophotal shapes as measured by the $a_4$ parameter
can either appear boxy ($a_4 <0$) or discy ($a_4 >0$). Boxy and discy galaxies seem to form a true dichotomy as the isophotal shape 
correlates with on first sight unconnected properties like X-ray luminosity \citep{1989A&A...217...35B} and central surface 
density slope (\citealp{1994AJ....108.1598F};\citealp{1995AJ....110.2622L};\citealp{1997AJ....114.1771F}). This led 
\citet{1996ApJ...464L.119K} to propose a revision of the Hubble sequence using the isophotal shape to cast elliptical 
galaxies into boxy and discy ellipticals, which might be a classification more closely connected to their formation 
history than apparent ellipticities \citep{2003ApJ...597..893N}.\\

Kinematic properties are normally measured from long-slit data at the major axis ($v$) and a 
central aperture ($\sigma$). But these are very special positions, which might or might not be representative
for the overall structure of the galaxy. It is now possible to extract the line-of-sight velocity distributions 
(LOSVDs) from the full two-dimensional field of a galaxy with modern integral-field instruments. The SAURON 
survey pioneered the first comprehensive sample of nearby E/S0 galaxies \citep{2002MNRAS.329..513D} with two-dimensional
kinematic data. Some of these galaxies exhibit very complex velocity or velocity dispersion fields which are not
easily captured by the usual $v/\sigma,\epsilon$ parameterization \citep{2004MNRAS.352..721E}. It can happen 
that two galaxies with similar long-slit measurements of ellipticity and $v/\sigma$ have morphologically 
different kinematic fields, e.g. one galaxy has a kinematically decoupled component and the other a regular 
velocity field. \citet{2007MNRAS.379..401E}, henceforth EM07, therefore introduced a luminosity weighted 
measure of the specific line-of-sight angular momentum  
\begin{equation}
\lambda_R=\frac{<R|V|>}{<R\sqrt{V^2+\sigma^2}>}, 
\end{equation}
where the pointed brackets are luminosity averages. $\lambda_R$ is designed to account for radial changes of angular 
momentum content in kinematic fields. This is different to the measure $<V>/<\sigma>$ proposed by \citet{2005MNRAS.363..937B} 
to exploit integral field kinematic data. However, \citet{2005MNRAS.363..937B} also stressed that with 
integral field data available, the connection between internal and projected structure of a galaxy is much less ambiguous.

The E/S0 galaxies are cast into the category of slow rotator or fast rotator according to their measured 
$\lambda_R$ (EM07): slow rotators have a $\lambda_R < 0.1$ and fast rotators $\lambda_R > 0.1$.
\citet{2007MNRAS.379..418C} showed that the slow and fast rotating galaxies of the SAURON sample populate distinct regions 
in the $(v/\sigma,\epsilon)$ diagram, even if their inclination is taken into account, which means that those galaxies
are also intrinsically fast or slowly rotating. 
Slow rotators tend to be more massive, have lower ellipticities and exhibit faster rotating, kpc-sized, old kinematically decoupled 
components (KDCs), while fast rotators have lower luminosities and can also have central KDCs, which are however small 
and often consists of young stellar populations \citep{2006MNRAS.373..906M}. Slow rotators also have strong kinematic twists 
and kinematic misalignments, while photometry and velocity contours are almost always aligned in fast rotators.
In addition, a correlation between anisotropy and ellipticity has been detected for slow and fast rotators
with slow rotators being more isotropic and fast rotators being anisotropic (\citealp{2007MNRAS.379..418C}; \citealp{2007arXiv0710.0663B}).

In this paper we want to connect the findings of the SAURON galaxies with their formation history. Ever since the work of 
\citet{1972ApJ...178..623T} the merger of two late-type galaxies has been so far the most appealing and the best tested
theoretical model for the formation of an elliptical galaxy (\citealp{1992ApJ...393..484B}; \citealp{1996ApJ...471..115B};
 \citealp{2003ApJ...597..893N}; \citealp{2006ApJ...650..791C}; \citealp{2006ApJ...641...21R}; \citealp{2006MNRAS.372..839N} and 
references therein).

The full two-dimensional kinematic field of merger remnants have scarcely been studied.
Following the pioneering work of \citet{2000MNRAS.316..315B}, \citet{2007MNRAS.376..997J}, henceforth J07, 
analyzed kinematic maps of  N-body merger remnants with kinemetric methods, introduced by \citet{2006MNRAS.366..787K}. 
They quantified the presence of counter-rotating cores, the flattening of isovelocity contours, 
counter-rotating discs and other peculiar features which are also observed in the SAURON galaxies. In 
general, maps of slow rotators resembled more equal-mass mergers while
fast rotators show kinematical fields similar to unequal-mass mergers. However, with the 
introduction of $\lambda_R$, we can quantify the line-of-sight angular momentum content of N-body 
merger remnants in the same way as has been done with observed galaxies. Ideally we can connect merger 
remnants which appear as slow rotator or fast rotator with a certain formation 
mechanism. Beyond that we know the true angular momentum content of a simulated N-body merger remnant and can 
therefore determine whether the remnant is indeed intrinsically a slow (fast) rotator or not. 

The aim of this work is two-fold: we want to study how much can be learned from the intrinsic structure of a 
galaxy by calculating its line-of-sight angular momentum parameter $\lambda_R$. In addition, we want to 
clarify whether remnants of binary 
mergers follow the slow-fast rotator division of the red sequence as found in the SAURON sample.
The paper is organized as follows: we describe the simulations in Sec.\ref{sec:sims}, give details about 
how we calculate the maps and determine $\lambda_R$ (Sec.\ref{sec:maps}) and examine 
the connection between line-of-sight angular momentum and intrinsic angular momentum (Sec.\ref{sec:angmom}), show
the projectional behaviour of $\lambda_R$ for merger remnants with varying orbital content (Sec.\ref{sec:shape}),
classify the maps derived from the merger remnants as fast or slow rotators and investigate the influence of
various formation mechanisms (Sec.\ref{sec:slowfast}) and finally discuss the results in Sec.\ref{sec:discus}. 
\section{Simulations}
\label{sec:sims}
\subsection{The Merger Models}
\label{MODELS}
The model parameters of the sample of 1:1 and 3:1 disc mergers we are studying
in this paper is identical to the sample presented in NJB2006, except
that we include star formation and stellar feedback in the modeling of the
dissipative component. Here we give a brief review of the setup.  

The disc galaxies were constructed in dynamical equilibrium using the
method described by \citet{1993ApJS...86..389H} with the following system of 
units: gravitational constant G=1, exponential scale length of the
larger progenitor disc $h_d=1$ (the scale height was $h_z=0.2$) and
mass of the larger disc $M_d=1$. The discs were exponential with an
additional spherical, non-rotating bulge with mass $M_b = 1/3$, a
Hernquist density profile \citep{1990ApJ...356..359H} and a scale
length $r_b=0.2h_d$, and a pseudo-isothermal halo with a mass $M_d=5.8$,
cut-off radius $r_c=10h_d$ and core radius $\gamma=1h$. The parameters
for the individual components were the same as for the collisionless
mergers presented in NB03 and NJB2006. We replaced 10\% of the stellar 
disc by gas with the same scale length and an initial scale height of 
$h_{z,gas} = 0.1 h_z$. 

Whereas the NB2003 and NJB2006 simulations had been performed with 
the Nbody-SPH code VINE \citep{2008arXiv0802.4253N,2008arXiv0802.4245W} 
we used the parallel TreeSPH-code GADGET-2 \citep{2005MNRAS.364.1105S} 
to follow radiative cooling \citep{1996ApJS..105...19K} for a primordial
mixture of hydrogen and helium together with a spatially uniform
time-independent local UV background \citep{1996ApJ...461...20H}. In addition, 
we include star formation and the associated supernova feedback following 
the sub-resolution multiphase model developed by \citet{2003MNRAS.339..289S} 
using the same implementation as in \citet{2008arXiv0802.0210J} but without feedback from 
black holes.   
 
We followed mergers of discs with mass ratios of 1:1 and 3:1. 
The equal-mass mergers were calculated using in total 440000 particles 
with each galaxy consisting of 20000 bulge particles, 60000 stellar 
disc particles, 20000 SPH particles representing the gas component in the disc, and 120000 halo
particles.  Twice as many halo particles than disc particles  were used 
to reduce heating and instability effects in the disc
components \citep{1999ApJ...523L.133N} by encounters between halo
and disc particles. For 3:1 mergers the parameters of the more 
massive galaxy were as described above. The low-mass companion contained a third the mass and
the number of particles in each component, with a disc scale-length
(stars and gas) of $h=\sqrt{1/3 }$. 

The gravitational forces were softened with a Spline kernel of $h_{\mathrm{grav}} = 0.05$. 
The minimal size of the Spline kernel used for computing the SPH
properties, $h_{\mathrm{SPH}}$, was fixed to the same
value. Implicitly, this procedure suppressed gas collapse on scales
smaller than the softening scale and prevents numerical instabilities
\citep{1997MNRAS.288.1060B}. 
The initial discs were run in isolation for two dynamical times to allow 
the systems to finally settle into an equilibrium state. The galaxies merged 
on parabolic orbits with a pericenter distance of 2 disc scale lengths. 
The initial geometries are identical to NB2003, Table 1.  and NJB2006. The merger remnants were  
allowed to settle into dynamical equilibrium for approximately 30 dynamical 
time-scales after the merger was complete. Then their equilibrium state was analyzed.    
All simulation have been run on a SGI Altix 
at the University Observatory in Munich. 

\subsection{Determination of $\lambda_R$}
\label{sec:maps}
The two-dimensional maps examined in this article, are identically determined
as laid out in paper J07. We repeat here briefly the most important steps: 
For the 2D analysis we binned particles within the central 6 length units (21 kpc) on a
grid of $48 \times 48$ cells. This corresponds typically to 2-3 effective 
radii depending on projection (see \citealp{2006MNRAS.tmp..463N} for the exact 
determination of $r_e$). To include seeing effects we created
for every luminous particle $10\times 10$ pseudo-particles with
identical velocities on a regular grid  with a total size
of 0.125 unit lengths (0.44 kpc) centered on the original particle
position. The mass of the original particle was then distributed to 
the pseudo-particles weighted by a Gaussian with a standard deviation 
of 0.1625 unit lengths (0.56 kpc). Thereafter the pseudo-particles were binned 
on a $48\times48$ grid.  
For the kinematic analysis we binned (mass weighted) all pseudo-particles 
falling within each grid cell in velocity along the line-of-sight. The width 
of the velocity bins was set to a value of 0.1 (26.2 km/s) for line-of-sight velocities
$v_{\mathrm{los}}$ in the range $ -4 \le v_{\mathrm{los}} \le 4 $ (corresponding to $\pm 1048$ km/s). 
This resulted in 80 velocity bins over the whole velocity interval. Using
the binned velocity data we constructed line-of-sight velocity
profiles (LOSVD) for each bin of the 2D grid. Subsequently we parameterized deviations 
from the Gaussian shape of the velocity profile using Gauss-Hermite basis functions
\citep{1993MNRAS.265..213G,1993ApJ...407..525V}. The kinematic parameters of 
each profile ($\sigma_{\mathrm{fit}}$, $v_{\mathrm{fit}}$, $h_3$, $h_4$) 
were then determined by least squares fitting. \\
$\lambda_R$ can be calculated from integrated field data via the formula given 
in EM07 
\begin{equation}
\lambda_R=\frac{\Sigma_{i=1}^{N_p} F_i R_i |V_i|}{\Sigma_{i=1}^{N_p}F_i R_i \sqrt{V_i^2 +\sigma_i^2}},
\end{equation}

where $F_i$ is the flux,$R_i$ the projected radius, $V_i$ the line-of-sight velocity and 
$\sigma_i$ the line-of-sight velocity dispersion of each grid cell. We determined these properties as explained
in the previous section and can therefore calculate $\lambda_R$ directly from our 2D data. As galaxies are 
projected randomly on the sky, we project each remnant 200 times randomly on the sphere of viewing directions. This 
gives us a sample of 20000 2-dimensional maps. SAURON has a finite field-of-view which extends to one 
effective radius, which we need to take  into account as $\lambda_r$ can vary strongly with R. We determine for 
each projection the effective radius and sum only over the grid cells inside one effective radius. Thus we ensure
 a fair comparison to the SAURON data.
\section{The Intrinsic Specific Angular Momentum and $\lambda_R$}
\label{sec:angmom}
\subsection{Intrinsic Angular Momentum}
The total intrinsic specific angular momentum $\mathrm{J_{intr}}$ of a galaxy (or N-body merger remnant) 
can be calculated from the discrete stellar (particle) distribution. The components of $\mathbf{J}$ in Cartesian 
coordinates are
\begin{equation}
\mathrm{J}_x= \sum_i y_i v_{z,i} - z_i v_{y,i} 
\label{eq:vec1}
\end{equation}
\begin{equation}
\mathrm{J}_y= \sum_i z_i v_{x,i} - x_i v_{z,i}
\label{eq:vec2}
\end{equation}
\begin{equation}
\mathrm{J}_z=\sum_i x_i v_{y,i} - y_i v_{x,i}
\label{eq:vec3}
\end{equation}
and $\mathrm{J}_{\mathrm{intr}}$ is the norm of that vector. For simplicity we implied here that all particles (stars) have 
the same mass and do not write the sums over masses explicitly. The angular momentum vector can projected to the plane of 
the sky, defining the {\it projected} intrinsic angular momentum vector $\mathbf{J_P}$ with components
\begin{equation}
\mathrm{J}_x= \sum_i y_i v_{z,i} - z_i v_{y,i}  
\end{equation}
\begin{equation}
\mathrm{J}_y=0 
\end{equation}
\begin{equation}
\mathrm{J}_z=\sum_i x_i v_{y,i} - y_i v_{x,i}
\end{equation}
where we define the y-direction arbitrarily as perpendicular to the plane of the sky. As our line-of-sight
is along the y-direction we can not observe transversal motions, i.e. in x and z-direction. We can now define
an apparent angular momentum vector $\mathbf{J_{app}}$ with the components
\begin{equation}
\mathrm{J}_x= \sum_i -z_i v_{y,i},
\end{equation}
\begin{equation}
\mathrm{J}_y=0, 
\end{equation}
\begin{equation}
\mathrm{J}_z=\sum_i x_i v_{y,i}.
\end{equation}
It has been show by \citet{1988skeg.book.....F} that for triaxial galaxies the apparent
angular momentum amounts to half the intrinsic angular momentum, which we rewrite as
\begin{equation}
\label{eq:franx}
\mathbf{J_{P}}=\mathrm{\kappa_J} \mathbf{J_{app}},
\end{equation}
where $\kappa_J$ is typically 2. An additional complication is that the angular momentum of a 
galaxy can be due both to streaming motions and to figure rotation. Again we follow \citet{1988skeg.book.....F}
\begin{equation}
\mathbf{J_{app}} = 0.5 \mathbf{J_{p}} + \mathrm{I_{proj}} \omega_\mathrm{proj},
\end{equation}
where $\mathrm{I_{proj}}$ and $\mathrm{\omega_{proj}}$ are the inertial moment tensor and the angular velocity 
of the projected density distribution, respectively. 
Indeed figure rotation is present in about 15\% of the merger remnants, but we do not try here to disentangle 
the relative contributions of figure rotation and streaming motion. Although such a decomposition would be 
possible in principle for the N-body remnants such an analysis would be beyond the scope of this paper. 
\subsection{Angular Momentum in Observations}  
\begin{figure}
\begin{center}
\epsfig{file=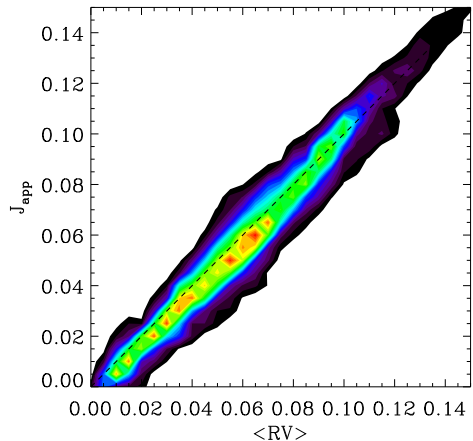,width=0.4 \textwidth}
\end{center}
\caption{The apparent angular momentum calculated from particle data (y-axis) as compared to the $\mathrm{<R|V|>}$ calculated 
from the velocity maps (x-axis). The agreement is very good for all projections and all merger remnants. \label{fig:jproj}}
\end{figure}

We cannot directly access the apparent angular momentum of a galaxy observationally. The closest proxy to 
the apparent angular momentum which we can calculate from the two-dimensional velocity field  
is $\mathrm{<R|V|>}$, where R is the projected radius, V the line-of-sight-velocity  and
brackets denote luminosity averages (see also Appendix A of EM07 for a discussion). It is 
straightforward to calculate $\mathrm{J_{app}}$ from the N-body simulation and $\mathrm{<R|V|>}$ from mock
observations of the same remnants and same aperture. In Fig.\ref{fig:jproj} we show that agreement 
is very good and verifies that the mapping of the velocity fields gives a close account of the 
intrinsic structure.

\begin{figure*}
\begin{center}
\epsfig{file=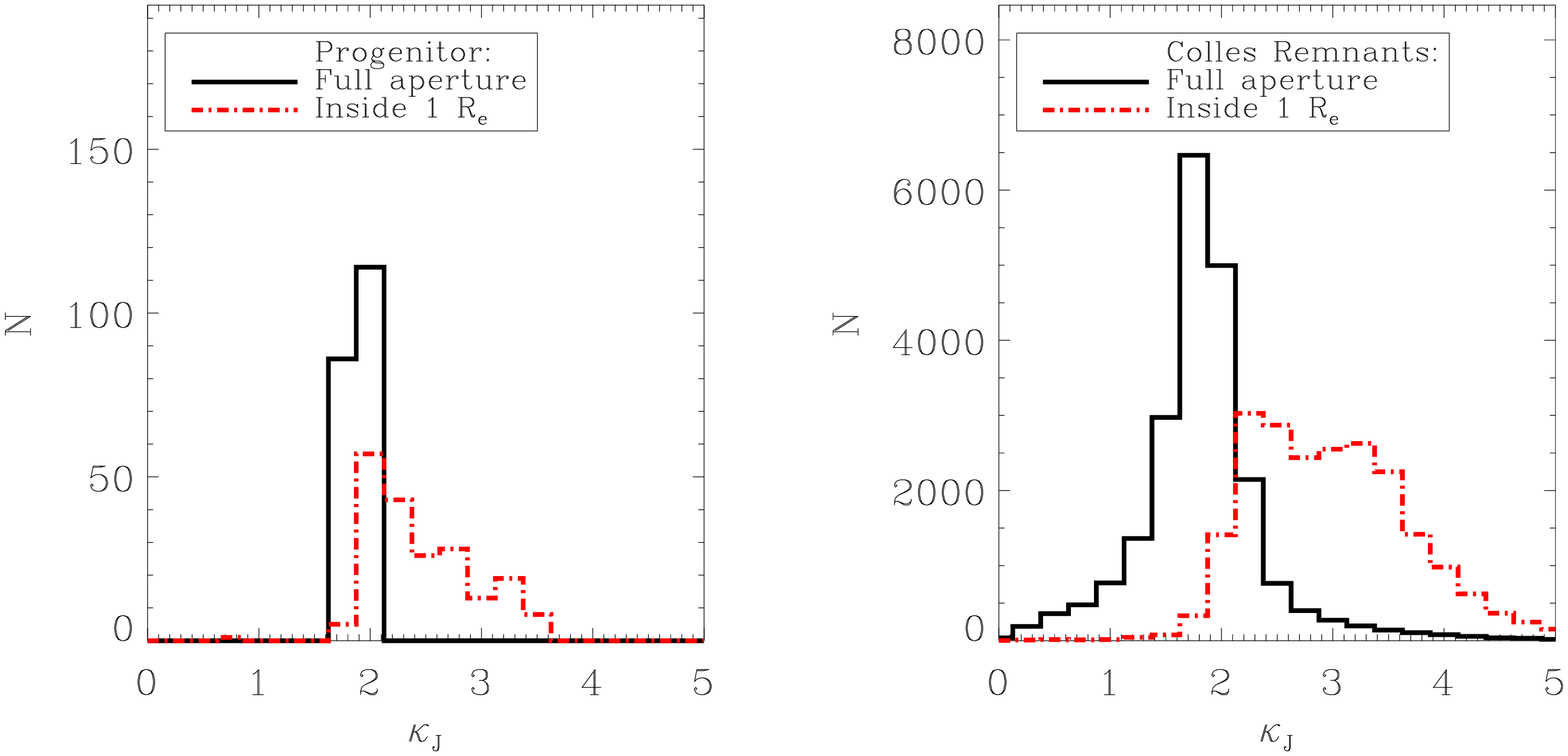,width=0.6 \textwidth }
\epsfig{file=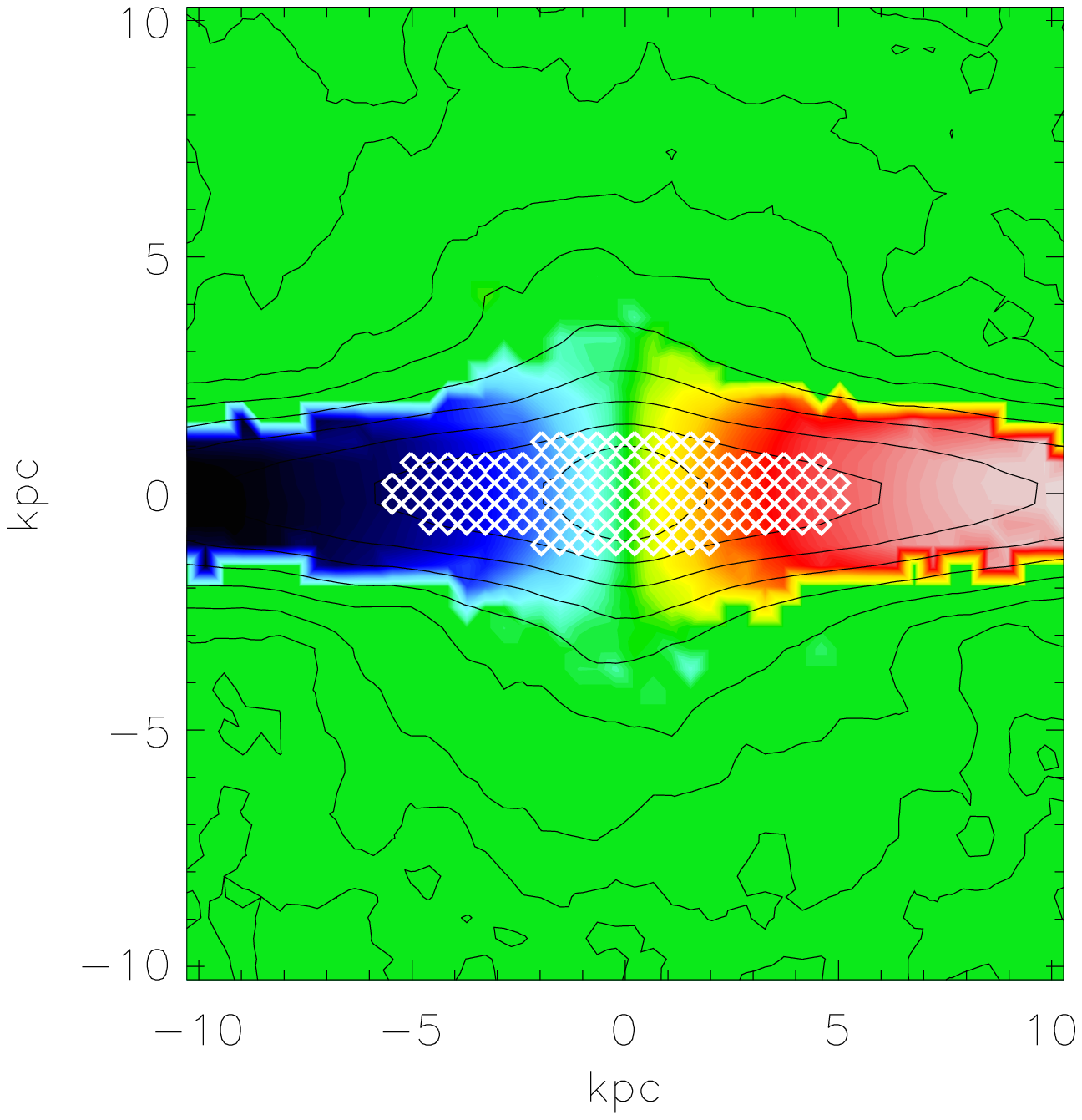,width=0.3 \textwidth}
\end{center}
\caption{Left: Distribution of the ratio $\kappa_J=\mathrm{J_{P}}/\mathrm{J_{app}}$ for the axisymmetric 
progenitor for 200 projections. The ratio shows the theoretically expected value of 2, if we calculate the angular 
momentum terms for the whole remnant, while a larger spread of $\kappa_J$ is observed if we constrain the field-of-view 
to one $R_e$. Middle: The same plot but this time for all collisionless remnants. Right: illustration of the size of the 
aperture at one effective radius (white rhombus). \label{fig:kappaj}} 
\end{figure*}

The inclination of a galaxy is in general not known and dynamical modeling or other observational
clues are needed to infer its projection \citep{2008arXiv0806.0042C}. We can choose the projection of an N-body remnant at will 
and can calculate $\mathrm{J_{app}}$ and $\mathrm{J_{P}}$ for each projection. The ratio of these two values should be 
equal to two, if and only if we have the information over the whole field of the galaxy. This is in general not the case, as the 
field-of-view is limited by the instrument we are using. We are calculating therefore $\mathrm{J_{app}}$ and 
$\mathrm{J_{P}}$ for two different apertures, one which encloses the whole remnant and one which extends only to 
one effective radius. This is illustrated in the right plot of Fig.\ref{fig:kappaj} where we highlighted the aperture
at one effective radius. 

It is instructive to examine first a test model, for which we chose the progenitor disc galaxy, which
is by construction axisymmetric and fast rotating. For almost all projections we find  that $\kappa_\mathrm{J}$ 
(Fig.\ref{fig:kappaj}, left) is very close to the theoretical value of 2, if calculated over the whole remnant. If we determine
$\mathrm{J_P}$ and $\mathrm{J_{app}}$ inside one $R_e$ we tend to get a spread of higher values for $\kappa_\mathrm{J}$. This is not too 
surprising, as we catch the particles only on part of their trajectory (close to pericenter) and line-of-sight velocities are not 
representative of their total angular momentum content. This will depend on the detailed orbit structure of the remnant, i.e. 
how much angular momentum sits at large radii. In the same Fig.\ref{fig:kappaj} we illustrate the spread for the
whole remnant sample, which is somewhat larger than for our axisymmetric test models, but with the same tendency  
to underestimate $\mathrm{J_P}$. Again using the information on the full remnants shows that the apparent angular momentum and the 
projected intrinsic angular momentum are on average connected by a factor of two. 

\begin{figure*}
\begin{center}
\epsfig{file=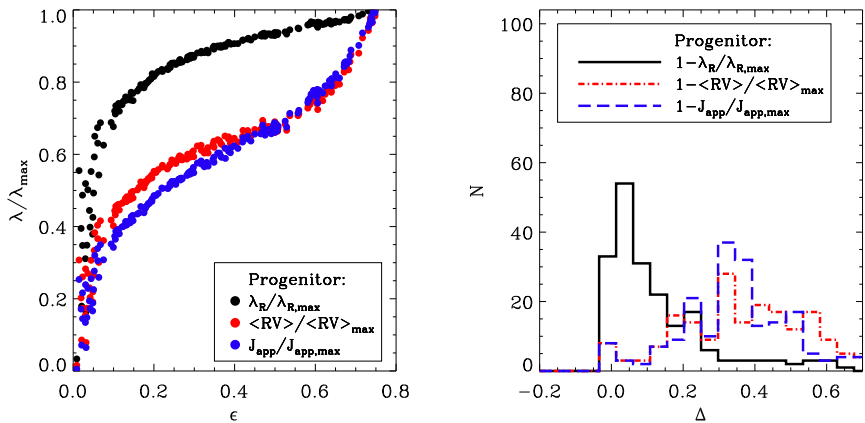,width=0.7 \textwidth}
\caption{Left: $\lambda_R$, $\mathrm{<R|V|>}$ and $\mathrm{J_{app}}$ for random projections of the progenitor galaxy. 
All quantities are normalized to their maximum value. Right: Distribution of absolute deviation of the same quantities 
from their (normalized) maximum values. \label{fig:jintr}} 
\end{center}
\end{figure*}

The quantity $\lambda_R$, as defined by EM07, has been proposed as a proxy for the intrinsic 
specific angular momentum of a galaxy within the radial range over which $\lambda_R$ is determined. 
In the last test we want to see how closely $\lambda_R$ and the angular momentum content are 
connected for a random projection. We discussed before that $\mathrm{<R|V|>}$, as a proxy 
for $\mathrm{J_{app}}$, is closely connected to the projected intrinsic angular momentum.  $\lambda_R$ is in 
contrast to $\mathrm{<R|V|>}$ a normalized quantity. Therefore we scale both quantities to their 
maximal values in order to compare the effect of inclination. We use again the progenitor disc galaxy as a test model 
for which ellipticity maps inclination accurately. We see in Fig.\ref{fig:jintr} that $\lambda_R$ decreases much 
less rapidly with inclination than $\mathrm{<R|V|>}$ does, which is also shown by the histograms of the deviations from their
maximal values (in this case scaled to one). The reason is that $<v>$ and $<\sigma>$ decrease 
simultaneously and their ratio stays roughly constant until we reach inclinations close to face-on. The true angular momentum 
content is of course most closely related to the maximum value of $\lambda_R$ or $\mathrm{<R|V|>}$, which in many cases, at least for 
fast rotators is a projection close to edge-on. In summary, even if there is a close connection between $\mathrm{<R|V|>}$ 
and the projected intrinsic angular momentum $\mathrm{J_P}$, the projectional behaviour of $\lambda_R$ 
is more robust and will stay closer to its maximal value for a higher fraction of all possible
viewing angles.

\section{Intrinsic Shape Variations: Example Remnants}
\label{sec:shape}
Before we examine the properties of the merger remnant sample as a whole it is instructive
to assess the variety of remnant properties present in the sample. 
The remnants in our sample have very different orbital makeups 
and intrinsic shapes. They can be dominated by major axis tubes, minor axis tubes 
and box orbits, and have prolate, oblate or triaxial shapes respectively \citep{2005MNRAS.360.1185J}. 
Their projected properties must correlate with their intrinsic shape and it is interesting to study
how  $\lambda_R$ is varying with inclination. In Thomas et al. 2007 we selected 
merger remnants at the extreme ends of the shape distribution and we want to 
study the projected 2-dimensional properties of three of them to elucidate systematic kinematical trends 
with intrinsic shape. 

\begin{figure*}
\epsfig{file=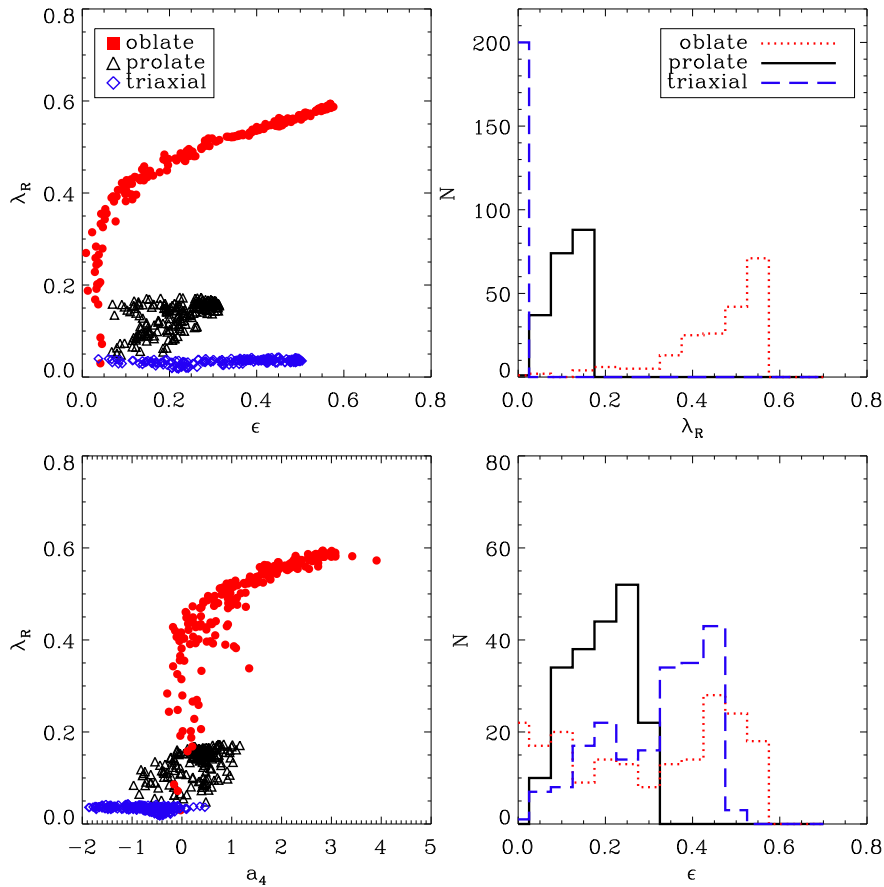,width=0.7 \textwidth}
\caption{Left: Case study of projectional relation between $\lambda_R$ and photometric properties, i.e. ellipticity and isophotal 
shape parameter for an oblate (red), a prolate (black) and a triaxial remnant (blue). The triaxial remnant has 
a $\lambda_R \approx 0$ not varying with projection. Ellipticity and $\lambda_R$ are good proxies for inclination angle for the oblate
remnant, which is nearly axisymmetric. The prolate remnant can have more than one $\lambda_R$ value for a given ellipticity 
owing to a complicated orbital structure. Right: Histogram of ellipticities and $\lambda_R$ of the projectional study. The oblate remnant 
is a fast rotator for almost all viewing angles, while the triaxial remnant is a slow rotator for all viewing angles. The prolate
remnant is equally probable to be observed as a fast or slow rotator. \label{fig:examples}} 
\end{figure*}

Our test remnants are the OBLATE, PROLATE and TRIAXIAL remnants from \citet{2007MNRAS.381.1672T}. 
We are plotting the ellipticity, $\lambda_R$ and the isophotal shape-parameter $a_4$ in Figure \ref{fig:examples}
to illustrate the variation with projection. The OBLATE remnant (red symbols) is the result
of a 4:1 merger, 
which is dominated by minor axis tubes. It is classified as a fast rotator from almost any 
viewing angle, although a dependence on inclination is visible. The TRIAXIAL remnant is a 1:1 merger remnant which is 
dominated by box and boxlet orbits. Interestingly the remnant has a $\lambda_R$ consistent with 0
from all viewing angles. The PROLATE remnant (black symbols) has the most complex 
dependency on inclination angle. This is so because it contains a significant amount of both minor and major axis tube 
orbits. Therefore we almost always see some rotation, originating from different kinds of orbit classes (for the detailed description
of the orbit classification in N-body remnants see \citet{2005MNRAS.360.1185J}. However, as this 
is also a 1:1 merger remnant it does not rotate very fast and reaches a maximum $\lambda_R$ of 0.18.

The statistical distribution of $\lambda_R$ and $\epsilon$ are better seen in the right column of Fig \ref{fig:examples}. 
The $\lambda_r$ of the TRIAXIAL remnant has a delta-function like distribution at $\lambda_R=0$, that means it is a
slow rotator under {\it any} viewing angle, while the OBLATE remnant is a fast rotator for almost all
projections except for three inclinations close to face-on. This results in a probability of $\approx 1 \%$ to 
misclassify the OBLATE remnant as a slow rotator. The PROLATE remnant is one of the few remnants 
which could be classified as a fast or a slow rotator with equal likelihood, because both tube orbit classes seem to carry 
an equal amount of rotation and move along the line-of-sight for different viewing angles.  
The PROLATE remnant is rather round for almost all projections, while the TRIAXIAL and the OBLATE remnant are quite
flattened, albeit the flattening  is caused by different orbit classes, i.e. box orbits in the case of the TRIAXIAL
remnant. Flattened slow rotators are observed in the SAURON sample only in exceptional cases (see also Sec.\ref{sec:slowfast}) 

\subsection{Kinematical Misalignment}
\begin{figure*}
\epsfig{file=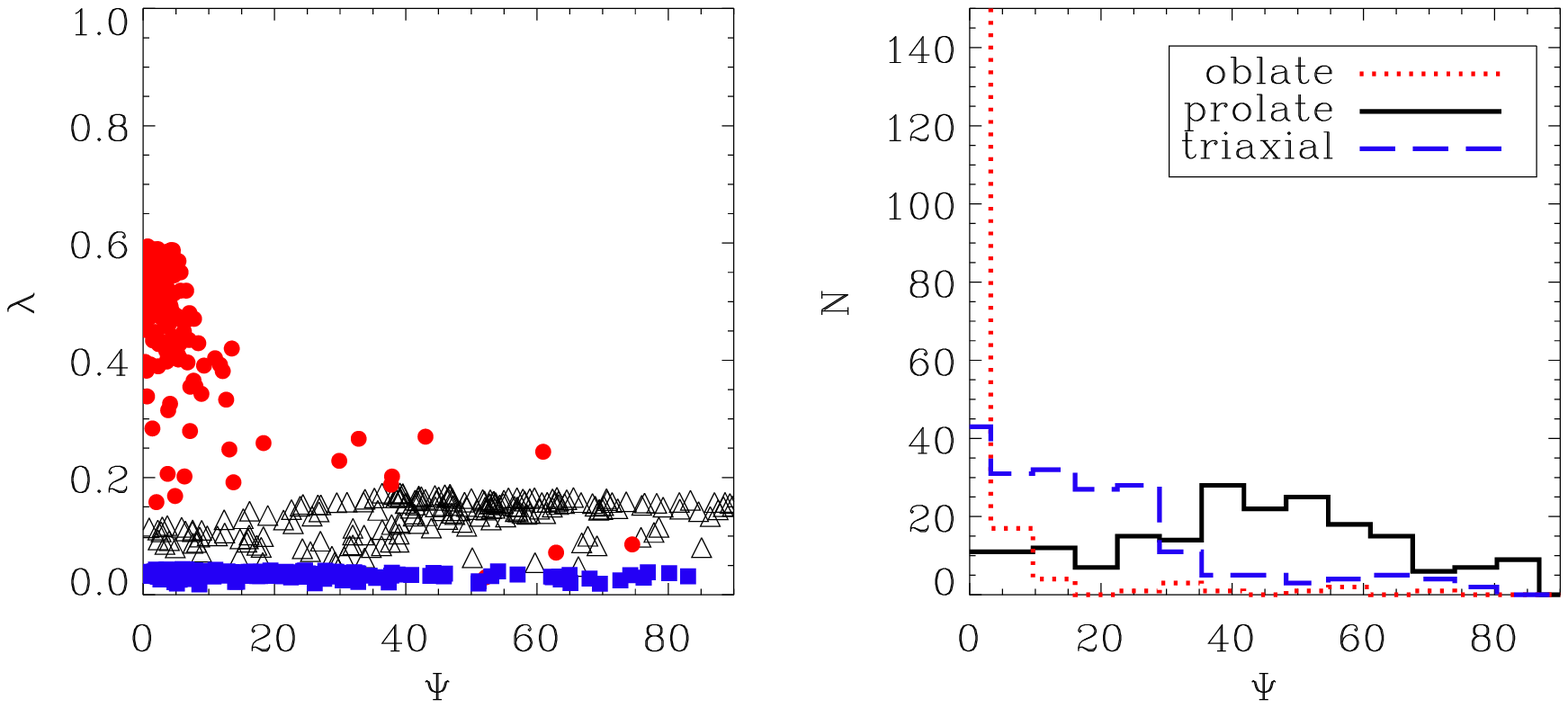,width=0.8 \textwidth}
\caption{Left: Case study of kinematical misalignment $\Psi$ of the same remnants as in Fig.\ref{fig:examples}. 
Relation between $\lambda_R$ and kinematic misalignment. Right: Distribution of kinematical misalignments from 200
random projections. The OBLATE remnant has very few projections with $\Psi > 5^\circ$, while the PROLATE remnant distribution
of misalignment angles peaks at $40^\circ$. The TRIAXIAL remnant has a less broad distribution of $\Psi$ values. 
 \label{fig:KM}} 
\end{figure*}

Another discriminatory property of slow and fast rotators is the kinematical misalignment. The misalignment angle 
$\Psi$ is defined as the angle difference between the photometric and the kinematic position angle
\begin{equation}
\Psi = |\mathrm{PA}_\mathrm{phot} - \mathrm{PA}_\mathrm{kin}|,
\end{equation}
where $\mathrm{PA}_{\mathrm{kin}}$ is defined as the angle $0^{\circ}< \mathrm{PA}_{\mathrm{kin}} < 180^{\circ}$ along 
which $|V|$ is maximal (see Appendix C of \citet{2006MNRAS.366..787K}). Triaxial systems can have complicated 
velocity fields, because of the possible superposition of orbits with angular momenta with respect to the 
minor and the major axis \citep{2008MNRAS.385..647V}. 
In a perfect axisymmetric galaxy there can be no kinematic misalignment. Indeed all fast rotators in the SAURON 
sample have misalignments $\Psi < 5^{\circ}$, whereas slow rotators can have any kinematic misalignment up to $90^{\circ}$. 
Our example remnants are also quite diverse in their behaviour of kinematic misalignment with projection angle (Fig.\ref{fig:KM}). 
The oblate remnant as it is dominated by minor axis tubes has very few projections where $\Psi \neq 0$. The triaxial
remnant has a much broader distribution, which peaks at low misalignments, but the maps of this remnant show little 
rotation, such that the misalignment has larger error bars and is not so meaningful. The prolate remnant, however, shows
clearly a strong misalignment, i.e. $\Psi > 20$, for most of the projections, as both tube orbit classes revolving around
the major as well as the minor axis, are present in this remnant.

\section{Slow and Fast Rotators}
\label{sec:slowfast}
The determination of the $\lambda$-parameter is important in sofar as ellipticals can be divided
into two broad sub-classes, so called slow and fast rotators (EM07). The (empirical) cut which separates
these classes has been made at $\lambda_R=0.1$. These seem to be genuinely intrinsically different galaxy 
types with little overlap through projection effects. Many slow rotators exhibit e.g. large, old and 
fast rotating kinematically decoupled components (KDCs), which however are not affecting too much the 
overall angular momentum balance, as the stars which move on high angular momentum orbits are not located in 
the center, such that the $\lambda_R$ as determined from the map is rather low. In general maps of slow 
rotators resemble more the kinematic maps derived from equal-mass merger remnants. The formation
of KDCs requires the inclusion of a gaseous component in the simulation. Fast rotators are 
on the other hand more likely to be un-equal mass mergers. For a detailed discussion on the origin 
of kinematical features in 2-dimensional LOSVD maps we refer the reader to J07.

\subsection{The Role of Merger Mass Ratios}
\label{sec:massratio}
\begin{figure}
\epsfig{file=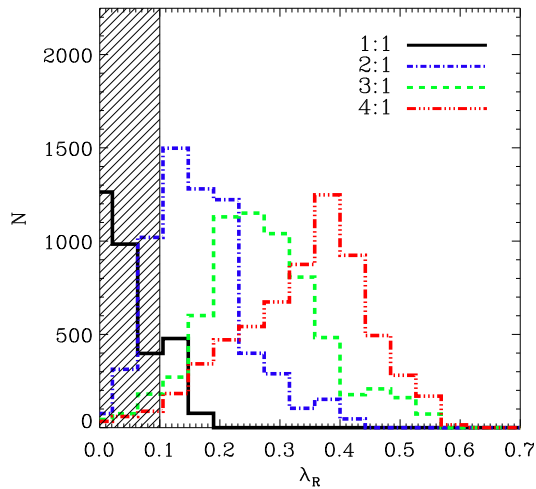,width=0.45 \textwidth}
\caption{Distribution of $\lambda_R$ for collisionless merger remnants of different merger mass ratios. The region
where kinematic maps are  classified as slow rotators is shaded. Only binary mergers of equal-mass progenitors 
seem to produce slow rotators in sizeable fractions. \label{fig:massratio} }
\end{figure}

The most fundamental parameter which influences the outcome of a merging event is the amplitude of the fluctuations
of the gravitational potential during the merger. How strongly the phase space is rearranged after a merger
 will be mainly determined by the ratio of the masses of the merging objects \citep{2003ApJ...597..893N}. An equal-mass 
merger will be more effective in redistributing the energies of the stars in the progenitor galaxies than an un-equal mass merger. 
It is true that secondary factors like infall velocity and merging geometry have also an impact on the final 
shape of the remnant, but the mass ratio of the merger is the dominant factor. 
We want to quantify when a given projection of a 
given remnant is classified as a slow or a fast rotator and we use collisionless disc-disc merger remnants 
of mergers with mass ratios of 1:1, 2:1, 3:1 and 4:1 from \citet{2003ApJ...597..893N}.
In Fig.\ref{fig:massratio} the $\lambda_R$ distribution is shown for each of these merger subsamples. We see, as expected, 
that the fraction of slow rotators for a given merger mass ratio is decreasing with increasing mass ratio. 
$\lambda_R$ peaks at 0, 0.1, 0.27 and 0.4 for the mass ratios 1:1 to 4:1. If one would assume that binary 
disc-merging is the main formation channel of elliptical galaxies then only equal or near equal-mass merger could produce 
them. However, this picture is probably too simple as elliptical galaxies can have more complicated merger histories in a cosmological
context (\citealp{2003ApJ...590..619M};\citealp{2007ApJ...658..710N}). Multiple minor mergers, e.g. three 3:1 mergers might 
also be a viable way to form slowly rotating ellipticals, which resemble closely the 
outcome of a single equal-mass merger, as was shown recently by \citet{2007A&A...476.1179B}. All the results (except
for Sec.\ref{sec:dry}) we present in this paper are valid for one generation of merging only. The examination of the impact of
complex hierarchical merging histories is beyond the scope of this paper. 
\begin{table}
\begin{center}
\begin{tabular}{cc}
\hline\hline
              Merger Mass Ratio                   & Slow Rotator Fraction  \\
\hline
1:1            & 0.75      \\
2:1            & 0.10      \\
3:1            & 0.03       \\
4:1            & 0.02      \\
\hline
\end{tabular}
\caption{Dependence of the slow rotator fraction on the merger mass ratio. Only equal-mass merger are very effective
in removing angular momentum in a single merging event. \label{tab:massratio} }
\end{center}
\end{table}
\subsection{Confusion Rate}
\begin{figure*}
\epsfig{file=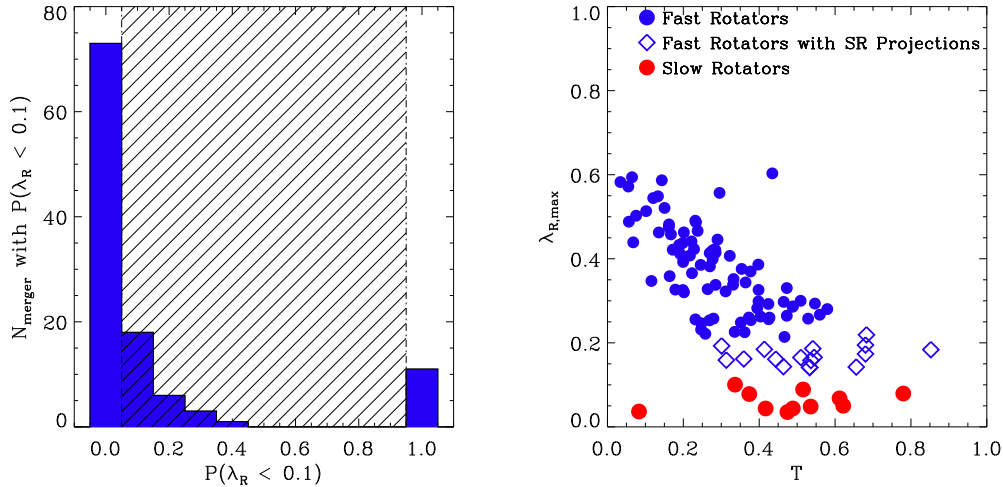,width=0.8 \textwidth}
\caption{Left: Distribution of mergers according to their probability of being classified
as a slow rotator ($\lambda_R < 0.1$). Right: Correlation between the maximal $\lambda_R$ and the 
intrinsic triaxiality T of the merger remnants. Open symbols indicate merger remnants which belong
to the shaded area from the plot on the left side, i.e. have a significant number of misclassified 
projections.  \label{fig:confuse}} 
\end{figure*}

\citet{2007MNRAS.379..418C} showed that slow and fast rotators are indeed intrinsic fast and slow rotators
and not randomly projected fast rotating galaxies which happen to appear non-rotating. We address this 
question in Fig. \ref{fig:confuse} where we examine the fraction of projections for each of the collisionless 
remnants which have $\lambda_R < 0.1$. There is an isolated sample of remnants which have a probability of greater
than 90\% to be classified as slow rotators. The remaining remnants have a probability of less than 40\% to 
be identified as slow rotators. The mergers which are located at the extreme ends of the distribution are the 
unambiguous cases they are (almost) always either fast rotators or slow rotators. 
There is however a small population of remnants which have a reasonable probability to be identified
as slow rotators although they have a non-zero angular momentum content (shaded zone in 
Fig.\ref{fig:confuse}, left). These are on average the most prolate of all remnants 
(same Figure, right plot). The confusion arises from the presence of major and minor axis tubes 
in the remnants, i.e. there are two face-on projections
under which the angular momentum content is perpendicular to the line-of-sight and cannot be observed.
But there also remnants which are mildly to strongly triaxial ($0.3 < T < 0.5$), which are hot but regularly 
rotating systems, i.e. with low kinematic misalignment. Almost all of these systems resulted from 
2:1 mergers. We can calculate now the global probability to falsely classify a rotating galaxy 
as a slow rotator in our sample and find $\mathrm{P_{confuse}}=0.046$. This is probably an upper limit as 
prolate remnants are overabundant in our sample, and are suppressed if a significant amount of 
gas is included in the simulations \citep{2006MNRAS.372..839N}. This is consistent with the SAURON sample
where not a single galaxy has been observed with rotational patterns as, e.g. our PROLATE model, which
we presented in Sec.\ref{sec:shape}. We further note that the maximal $\lambda_R$ is a good tracer 
for the triaxiality of our remnants (Fig.\ref{fig:confuse},right), albeit with an increasing spread 
towards lower $\lambda_R$. Particular spin alignments of the progenitor discs can lead to interesting 
outliers from this correlation, e.g. we find a slow rotator with $T \approx 0$. This merger remnant 
originated from a planar merger of discs with anti-aligned spins. The counter-rotation of both disc
components cause a net zero rotation. While this is a possible formation scenario, most slow rotators 
are triaxial box dominated remnants.  

\subsection{The Influence of Dissipation}
\begin{figure}
\begin{center}
\epsfig{file=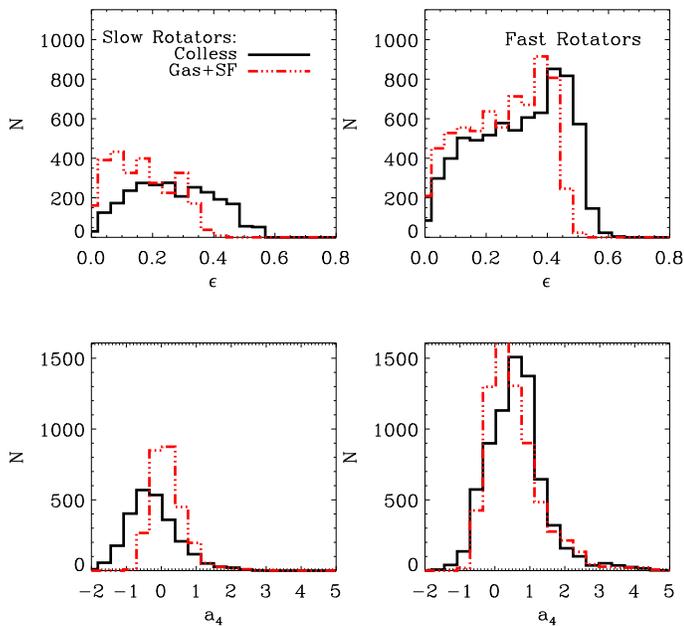,width=0.5 \textwidth}
\caption{Distribution of ellipticities and isophotal shape parameter for the slow rotator (left plots) 
and the fast rotator (right plots) population. The collisionless 1:1 and 3:1 remnants are indicated by 
black solid lines. Fast and slow rotators have clearly separate photometric
properties. Fast rotators have high ellipticities and discy isophotes. Histograms for dissipatively formed
merger remnants are overplotted in (dashed) red. Fast rotators are very little affected while slow rotators
with gas have significantly lower ellipticities, i.e. are rounder.  \label{fig-1} }
\end{center}
\end{figure}

\begin{figure*}
\epsfig{file=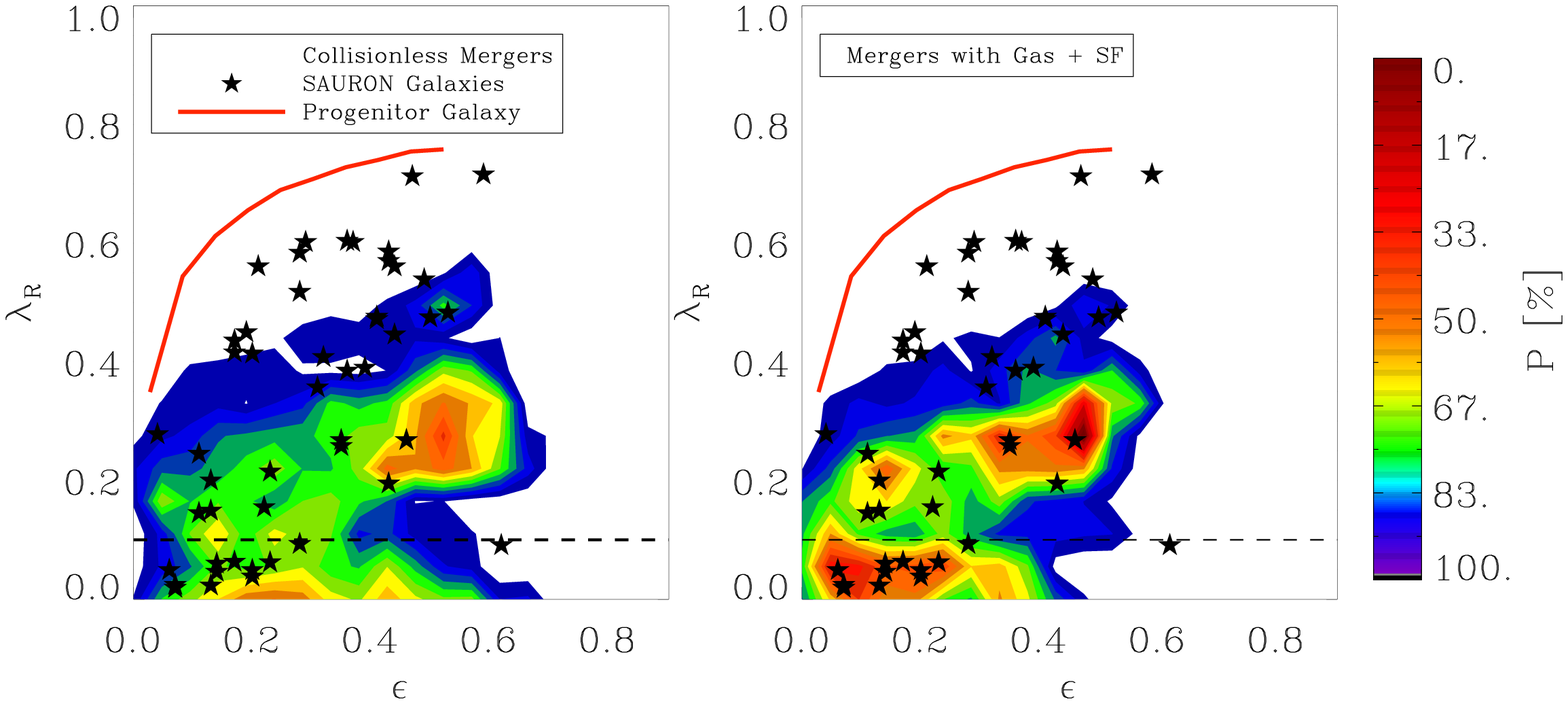,width=0.85 \textwidth }\\
\epsfig{file=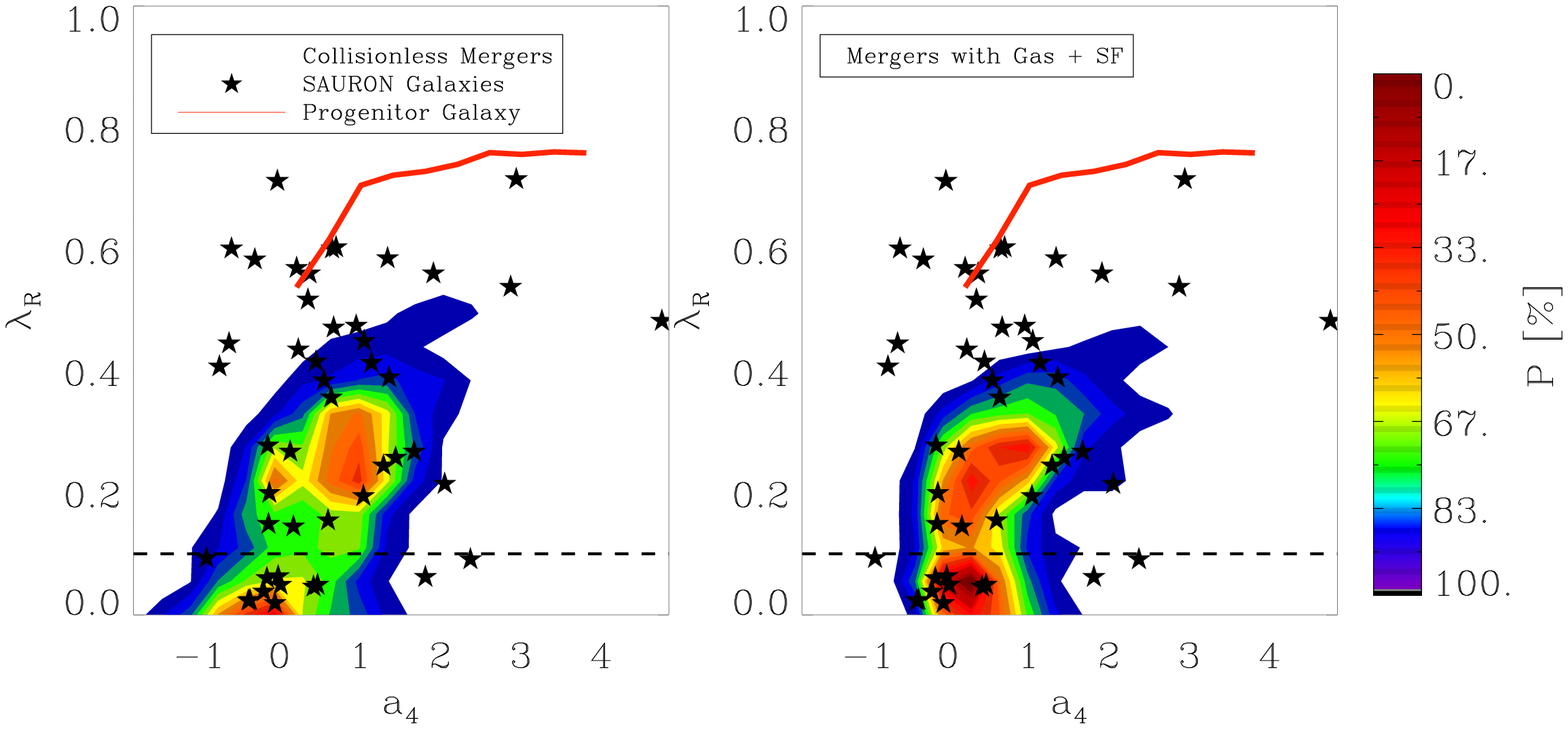,width=0.85 \textwidth }
\caption{Left Column: Two-dimensional probability distribution of collisionless 1:1 and 3:1 remnants in $\lambda_R - a_4$, 
respectively in the $\lambda_R - \epsilon$ plane. The SAURON galaxies are indicated by black stars. High 
$\lambda_R$ galaxies are resembling more the progenitor galaxy than 3:1 merger remnants. Boxy fast rotators
are not formed in mergers, but are probably barred galaxies. Right Column:
The same plots but this time with merger remnants which formed with gas. The photometric properties
are stronger affected than $\lambda_R$. Slow rotating remnants agree well with 
SAURON galaxies.\label{fig-2}}
\end{figure*}

Late-type spiral galaxies are normally not purely collisionless systems but have a sizable fraction of gas.
We therefore compare  two sets of mergers: disc-disc mergers  with gas and star formation, and collisionless 
disc-disc mergers. Each set consists of 16 1:1 and 32 3:1 mergers, which formed on identical merging geometries. As 
mentioned before we take 200 observations of each remnant at random viewing angles, which results in 9600 maps for each 
set of simulations. 
We determine for each set the fraction of slow and fast rotators, as well as standard photometric parameters such as
ellipticity and the isophotal shape parameter $a_4$. In Fig. \ref{fig-1} we show the distributions of ellipticities 
and $a_4$ for slow and fast rotators from mergers which formed with and without gas. Maps classified as slow and 
fast rotators show distinct photometric properties. Slow rotators in general have smaller ellipticities 
and have smaller $a_4$ values while fast rotators are more elliptical and discy.
It is apparent that the fast rotator population hardly changes its photometric properties when gas is 
included in the simulation.
The impact on slow rotators, however, is more visible as the ellipticity distribution is now more skewed towards $\epsilon=0$, 
while the $a_4$ distribution peaks now at 0 (\citealp{2006ApJ...650..791C};\citealp{2006MNRAS.372..839N}). 
The mean slow rotator properties are summarized in Table \ref{tab-1}, which also confirm that most 
slow rotators originate from equal-mass mergers. 
The results from the SAURON survey showed that slow and fast-rotating ellipticals are located at certain positions in 
the $\lambda -\epsilon$ and in the $\lambda - a_4$ plane. For a better comparison we 
also show the distribution of merger remnants in those parameter spaces (Fig. \ref{fig-2}). In general the parameter space 
bracketed by the SAURON observations is reproduced well by the merger remnants. There are, however, some differences.
For example
we see that the collisionless mergers have too high $\epsilon$ for a given $\lambda_R$ which is alleviated, especially 
for slow rotators, by the inclusion of gas. We also indicated the progenitor galaxy in Fig.\ref{fig-2}. We already showed 
in Sec.\ref{sec:massratio} that the amount of rotational support always decreases after a merger event. 
Very high merger mass ratios will lead to merger remnants which would lie closer to the progenitor limit, but we have not performed
10:1 or even higher mass ratios. Therefore we do not find projections of merger remnants with $\lambda_R > 0.6$. It is remarkable 
though that some of the SAURON fast rotators have ($\lambda_R$,$\epsilon$)-values very close to the progenitor galaxy 
and are, at least in this sense, hardly distinguishable from early-type spiral galaxies.

The situation is different in the $a_4 - \lambda$-plane where the merger remnants are only partly 
located in the parameter space occupied by the SAURON galaxies. Collisionless slow rotators are too boxy  while fast rotators 
are on average too discy compared to SAURON galaxies. While gas removes the boxy projections of slow rotators it does not help
to create, as expected, boxy projections in fast rotators. Probably the origin of boxiness in the
fast rotating SAURON galaxies is different from the source of boxiness in our remnants, e.g. bars.

\begin{figure}
\epsfig{file=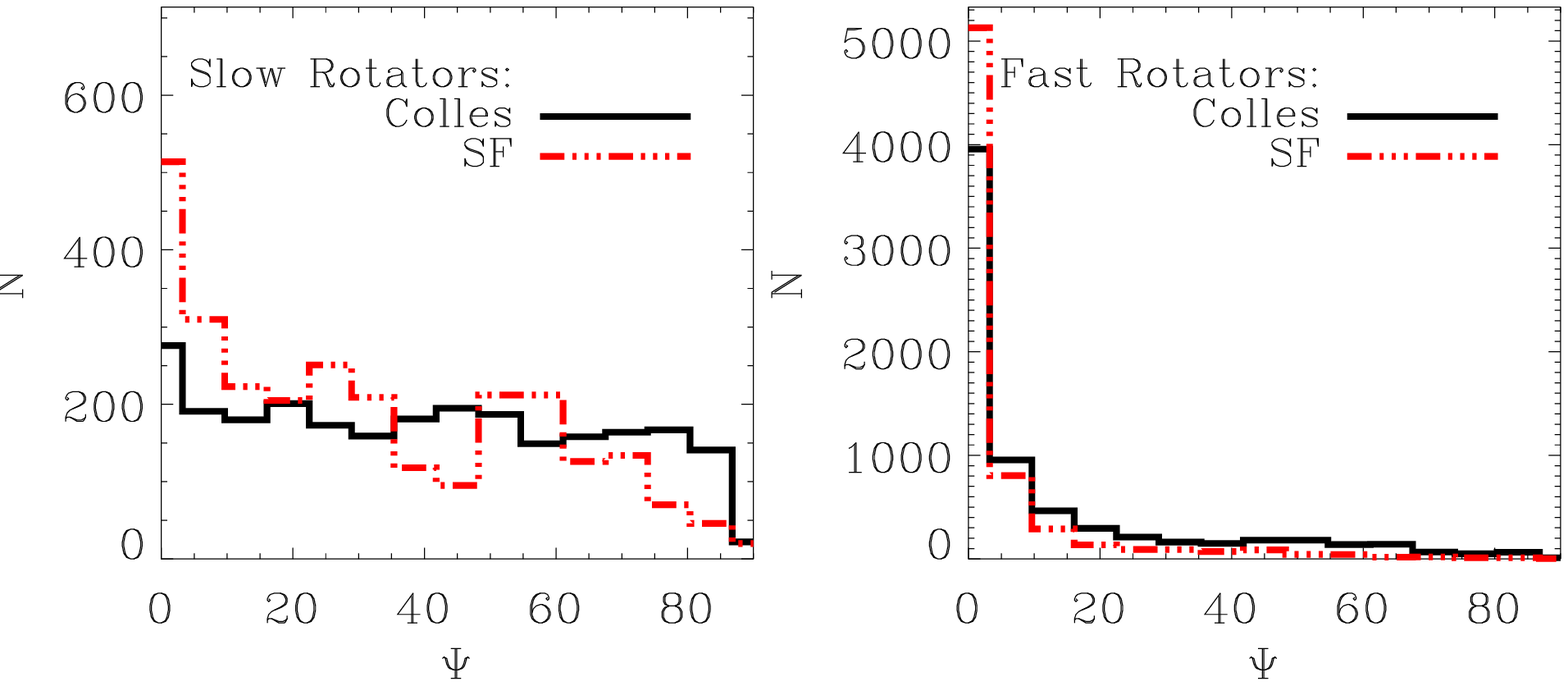,width=0.5 \textwidth }
\caption{Histograms of kinematic misalignment angle $\Psi$ measured from 2-dimensional kinematical maps 
of the merger remnants. The overwhelming majority of maps identified as fast rotators have no 
kinematic misalignment, while slow rotators have no preferred misalignment angle. Dissipation changes
the results only slightly, i.e. the maps slow rotators have appear slightly more aligned. \label{fig:kmhisto}}
\end{figure}

As we discussed earlier, kinematical misalignment is also a distinguishing feature between fast and slow rotators.
Again we divide up the simulation sample into fast and slow rotators and measure the angle $\Psi$ between kinematic
and photometric position angles. The observational trends are reproduced remarkably well: fast rotators 
are very unlikely to show kinematic misalignment, while slow rotators show a flat distribution of misalignment
angles, in other words there is no preferred alignment angle for slow rotators (see Fig.\ref{fig:kmhisto}). 
The low misalignment angles for fast rotators are not very surprising, because we already start out with 
a perfectly aligned system. Unequal mass merger simply do not destroy the initial alignment. Indeed observed fast 
rotators are also remarkably aligned. Only 2 out of 50 fast rotators have $\Psi > 5^\circ$ (see \citealp{2007MNRAS.379..418C}, Fig.12). The 
crucial question to explain the origin of fast rotators is rather how the axisymmetric progenitor systems are 
produced in the first place, rather than the subsequent merging.

Gas changes the picture only slightly. There are somewhat fewer projections with high misalignment angles for fast 
rotators and slow rotators are slightly better aligned. In general the inclusion of only 10\% of gas in the merging 
process is not necessary to explain the kinematic misalignment of slow and fast rotators. However, the inclusion of high 
fractions of gas would probably change this picture. 

\begin{table}
\begin{center}
\begin{tabular}{lcccccc}
\hline\hline
                                 & Collisionless   &  Gas + SF  \\
\hline
Slow Rotator Fraction            & 0.26     &  0.28    \\
Slow Rotator $<\epsilon>_{med}$   & 0.28     &  0.18    \\
Slow Rotator $\epsilon < 0.3$    & 0.55     &  0.81    \\
Slow Rotators from 1:1 merger    &  0.95    &  0.94   \\
\hline
\end{tabular}
\caption{Slow Rotator Properties. Main results of the classification according to the $\lambda_R$-parameter. Every
projection which has a $\lambda < 0.1$ is classified as a slow rotator. Dissipation mainly influences the ellipticity, while
the slow rotator fraction is almost unaffected. \label{tab-1} }
\end{center}
\end{table}

\subsection{Dry Mergers}
\label{sec:dry}
\begin{figure}
\epsfig{file=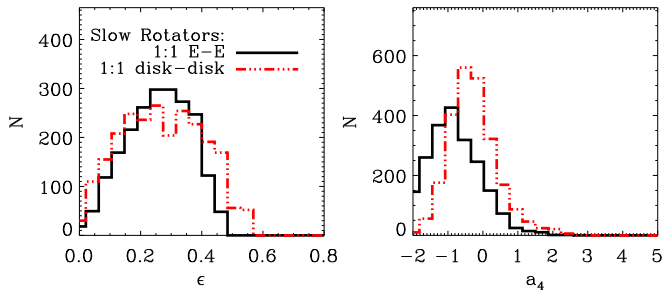,width=0.5 \textwidth}
\caption{Distribution of photometric properties for slow rotators from collisionless 1:1 elliptical and 1:1 disc mergers. 
The ellipticity distribution of both samples are similar while elliptical merger remnants are more boxy. \label{fig-dry} }
\end{figure}

The most massive galaxies in the universe are probably not formed from a single
generation of binary spiral-spiral mergers \citep{2007astro.ph..2535N}, because there are simply not enough massive late-type 
galaxies to account for the stellar mass of galaxies with luminosities, e.g., of 4 $L^*$. Their merging history included 
probably more than one major merger and semi-analytic modeling showed that the last major merger 
was probably dry (\citealp{2003ApJ...597L.117K};\citealp{2007ApJ...659..976H}). We have a small sample 
of re-mergers of collisionless merger remnants to test the assumption if slow rotators are the end product 
of dry merging. We analyzed 6 equal-mass mergers of discy remnants and 6 equal-mass mergers of boxy remnants. 
The majority of the maps (about 68 \%) of the dry mergers are indeed slow rotators. As we have a very limited 
sample of merger remnants, we compare them only with the slow rotators which originated from 1:1 collisionless 
disc mergers, which is admittedly the overwhelming majority. In Fig. \ref{fig-dry} we show  that the ellipticity 
distribution from elliptical-elliptical mergers is hardly distinguishable
from equal-mass disc-disc mergers. However, they are on average more boxy than remnants of  
disc mergers in agreement with \citet{2006ApJ...636L..81N}. The slow rotators in the SAURON sample are indeed
hardly boxy. As mentioned before, the SAURON sample is not representative of the shape distribution of all elliptical
galaxies due to the selection procedure of galaxies. There is especially a dearth of triaxial, boxy galaxies
which makes it difficult to exclude or confirm the dry merging origin of massive ellipticals with the present
sample (see also \citet{2007arXiv0710.0663B} for a more detailed discussion of the origin of the slowly rotating
SAURON ellipticals).

\section{Discussion and Conclusions}
\label{sec:discus}
We calculated the $\lambda_R$-parameter from line-of-sight velocity maps of 48 remnants of collisionless disc 
mergers, 48 remnants of disc mergers with a dissipative component and 
12 re-mergers of collisionless remnants, according to the procedure laid out in 
EM07. Every remnant is projected 200 times randomly and then treated as a mock galaxy.
We first investigated how the intrinsic properties of the merger remnants manifest itself in the maps and how much
can be deduced of the intrinsic structure by calculating $\lambda_R$.  We then identified a mock galaxy as a fast or 
slow rotator and compare their photometric and kinematic properties like $a_4$, $\epsilon$ and kinematic misalignment 
$\Psi$ to the properties of SAURON galaxies. We examined the influence of gas and merging mass ratio on the statistical
properties of slow and fast rotators. 
We report the following findings:  
\begin{itemize}
\item
The line-of-sight angular momentum parameter $\lambda_R$ is a rather robust indicator of the
intrinsic angular momentum content of a galaxy. $\lambda_R$ stays close to its maximum value for a wide range of 
viewing angles as it captures the balance between ordered (line-of-sigh velocity) and unordered motion
(line-of-sight velocity dispersion) well. In principle the projected angular momentum content could also be deduced from the 
two-dimensional velocity field alone, but information over the entire field of the galaxy would be needed, which is 
normally not available. We conducted therefore tests to explore the deviation from the intrinsic angular 
momentum, observing galaxies with apertures of limited size.
\item
The projectional behaviour of $\lambda_R$ is determined by the intrinsic orbital structure.
Most fast rotators consist of disc-like minor axis tubes and have $\lambda_R > 0.1$ for almost all projections.
Likewise slow rotators consist of box orbits and have $\lambda_R < 0.1$ for all projections. Prolate remnants have 
a more complicated projectional behaviour as they consist of major and minor axis tubes. There is however not a single 
galaxy in the SAURON survey with significant rotation around the major axis and we slightly overproduce these 
systems in our merger sample. Even so we determine the misclassification probability for the whole collisionless
merger sample to only 4.6 \%. We therefore confirm the robustness of the slow/fast rotator classification scheme 
for binary disc merger remnants. 
\item
The spread of $\lambda_R$ is naturally reproduced in the binary merger scenario. However their photometric
properties have some important differences. Slow rotators from collisionless mergers have too high ellipticities. The 
problem is alleviated by the inclusion of a modest fraction of gas, which does not modify the slow rotator fraction
but produces remnants with lower ellipticities. Therefore gas, maybe counter-intuitively, can be important for 
the formation of slow rotators.
\item
The destruction of ordered rotation is only effective in equal-mass mergers. The violence of the merger seems
to be the decisive factor, i.e. with increasing mass-ratio the percentage of formed slow rotators drops
very rapidly. This is true at least for the binary merger scenario, very large gas fractions and multiple mergers will
complicate this picture. 
\item
We tested the amount of kinematic misalignment for slow and fast rotators separately. Again the agreement
with observations is very good. Slow rotators can have any misalignment with equal probability, while fast rotators 
show kinematical alignment for almost all projections. The picture changes with the inclusion of gas only insofar as the
strongest misalignments vanish for slow rotators. Higher gas fraction will probably lead to a further decrease in observed
misalignments. 
\item 
Slow rotators in the SAURON sample are not strongly boxy and rather round, therefore the dry mergers 
are a poor fit to them. This is somewhat surprising, as the projection and mass ratio independence of
boxiness was seen as an asset of this formation mechanism \citep{2006ApJ...636L..81N}.   
To assess the role of dry mergers adequately probably a larger sample of ellipticals, like ATLAS$^{\rm 3D}$, is needed.
\item
Fast rotators with boxy isophotes are absent from our sample. Some early-type S0s in the SAURON
are barred galaxies, and such strong bars seem not to be formed in our particular merger sample. Our sample
of course has its limits and it is possible that the bulge to disc ratio of our progenitor galaxies is too high 
or the initial bulges are not rotating or probably both, to compare to this particular property of the SAURON galaxies.
\item 
The fast rotators with the highest $\lambda_R$ values in the SAURON sample resemble our progenitor galaxy more 
than a merger remnant, which leads to the question, if major merging played a significant role in their formation
at all. 
\end{itemize}

\section*{Acknowledgments}
This research was supported by the DFG priority program SPP 1177 and by the DFG 
cluster of excellence 'Origin and Structure of the Universe'. Part of the simulations
were run on the local SGI ALTIX 3700 Bx2 which was also partly funded by this cluster of excellence.
MC acknowledge support from a STFC Advanced Fellowship (PP/D005574/1).
\bibliographystyle{mn2e}
\bibliography{references}

\clearpage

\clearpage

%% Use the figure environment and \plotone or \plottwo to include
%% figures and captions in your electronic submission.
%% To embed the sample graphics in
%% the file, uncomment the \plotone, \plottwo, and
%% \includegraphics commands
%%
%% If you need a layout that cannot be achieved with \plotone or
%% \plottwo, you can invoke the graphicx package directly with the
%% \includegraphics command or use \plotfiddle. For more information,
%% please see the tutorial on "Using Electronic Art with AASTeX" in the
%% documentation section at the AASTeX Web site,
%% http://www.journals.uchicago.edu/AAS/AASTeX.
%%
%% The examples below also include sample markup for submission of
%% supplemental electronic materials. As always, be sure to check
%% the instructions to authors for the journal you are submitting to
%% for specific submissions guidelines as they vary from
%% journal to journal.

%% This example uses \plotone to include an EPS file scaled to
%% 80% of its natural size with \epsscale. Its caption
%% has been written to indicate that additional figure parts will be
%% available in the electronic journal.

%% This figure uses \includegraphics to scale and rotate the still frame
%% for an mpeg animation.

%\begin{figure}
%\includegraphics[angle=90,scale=.50]{f3.eps}
%\caption{}
%\end{figure}

\clearpage

%% If the table is more than one page long, the width of the table can vary
%% from page to page when the default \tablewidth is used, as below.  The
%% individual table widths for each page will be written to the log file; a
%% maximum tablewidth for the table can be computed from these values.
%% The \tablewidth argument can then be reset and the file reprocessed, so
%% that the table is of uniform width throughout. Try getting the widths
%% from the log file and changing the \tablewidth parameter to see how
%% adjusting this value affects table formatting.

%% The \dataset{} macro has also been applied to a few of the objects to
%% show how many observations can be tagged in a table.

\end{document}